# Crystal growth and characterization of $MgB_2$: Relation between structure and superconducting properties

S.Lee[1*], T.Masui[1,2], H.Mori[1], Yu.Eltsev[1], A.Yamamoto[1] and S.Tajima[1]

[1] Superconductivity Research Laboratory, ISTEC, 1-10-13 Shinonome, Koto-ku, Tokyo 135-0062, Japan

[2] Domestic Research Fellow, Japan Science and Technology Corporation

**Abstract**

We discuss the important aspects of synthesis and crystal growth of $MgB_2$ under high pressure (P) and temperature (T) in Mg-B-N system, including the optimisation of P-T conditions for reproducible crystal growth, the role of liquid phases in this process, the temperature dependence of crystal size and the effect of growing instabilities on single crystals morphology. Extensive experiments have been carried out on single crystals with slightly different lattice constants and defects concentration, which revealed and possible effects of Mg-deficiency and lattice strain on the superconducting properties of $MgB_2$ ($T_c$, $J_c$, residual resistivity ratio, anisotropy etc.).

*Keywords: $MgB_2$, single crystal growth, high-pressure technique, Mg-deficiency, lattice strain, magnetic and transport properties.*

[*] Corresponding author:

Dr. Sergey Lee

Superconductivity Research Laboratory, ISTEC,
1-10-13 Shinonome, Koto-ku, Tokyo 135-0062, Japan
Phone: +81-3-3536-0618;  Fax: +81-3-3536-5717;  E-mail:  lee@istec.or.jp

**Introduction**

The discovery of superconductivity in magnesium diboride ($MgB_2$) with $T_c$=39K [1], the highest value known for the non-cuprate phases, has renewed the interest to the simple non-oxide compounds and stimulated the worldwide research activity on its fundamental physical properties and various aspects of material fabrications. The simple chemical composition and crystal structure, the relatively high $T_c$ value and the "weak-link"-free grain boundaries make $MgB_2$ an attractive candidate for both of basic study and application [2].

Extensive theoretical and experimental studies performed during last one-and-half year showed that despite of the chemical and structural simplicity, the band structure of $MgB_2$ is not simple, characterized by the four bands that could give two superconducting gaps with different characters [3-6]. The layered crystal structure consisting of alternating layers of Mg and B atoms causes an appreciable anisotropy ($\gamma$) of superconducting parameters [7,8] that changes significantly with temperature [9-10].

Although $MgB_2$ can be synthesized by a straightforward method (from a low cost and nontoxic reagents) the synthesis of high-quality (phase pure) samples is still quite challenging. It requires a heat treatment of reactant mixture in a closed reaction vessels or utilization of external pressure of inert gases to prevent evaporation and oxygenation of magnesium. Different synthetic techniques and/or variation of processing parameters produce the polycrystalline samples with slightly different $T_c$, residual resistivity, flux pinning properties etc., which was attributed to the effect of impurities, structural defects (i.e. Mg-vacancies) and lattice strains [11-18]. In addition to these problems, the low chemical stability of $MgB_2$ causes degradation at ambient atmosphere. All these facts are considered to be the reasons for the discrepancies of experimental data reported for polycrystalline samples.

The growth of high-quality single crystals of sub-millimeter size has opened a route for the various physical studies involving the X-ray structural analysis [7,19], the measurements of magnetism [20-23], electric transport [8,24-26], thermal conductivity [9], Hall effect [27] penetration depth [28], de Haas van Alphen effect [6], Raman scattering [29], angle-resolved photoemission [4] and optical spectra [30], which are important for understanding of the fundamental properties of $MgB_2$ superconductor.

Two different methods were developed for the crystal growth of $MgB_2$. One is heat treatment in sealed metal (Mo, Nb, stainless steel) containers [31-33] and the other is the sintering at high-pressures and temperatures in BN [7,10,25] or Ta reaction cells [34]. Although most single crystal data are in reasonable agreement, the apparent differences of crystal size, morphology, structural and superconducting properties are reported for different methods by several research groups.

This paper describes our recent results on $MgB_2$ crystal growth under high pressure in Mg-B-N system and the effect of processing parameters on crystal growth, structural and superconducting properties of grown crystals.

## Experimental

Single crystals were prepared from magnesium powder (100 mesh, 99.9% Rare Metals) and amorphous boron (0.9 μm, 97% Hermann C.Starck). The 3% impurity in the boron reagent includes about 2% of moisture that was removed before crystal growth experiments by annealing in dynamic vacuum, and the major metal impurity was Mg (0.39 %). For the experiments in Mg-B-N system, the BN powder (99.99% High Purity Chemical Inc.) was used as a source of boron and nitrogen. Reaction cells utilized in this study were 93.2% (HC-type) and 99.2% (N-1 type) pure BN containers (Denki Kagaku Kogyo K.K.) with an inner diameter of 5 mm, a wall thickness of 1 mm and a length of 10.5 mm. The BN (HC) has a lower purity than BN (N-1) because of a higher concentration of CaO and $B_2O_3$ impurity phases.

High-pressure experiments were performed using cubic-anvil press (Try Engineering Co.) at a pressure (P) of 2-5 GPa. Synthesis of polycrystalline samples and crystal growth experiments were carried out in a wide temperature range of 900-1700°C. Owing to a longitudinal temperature gradient, the temperature difference (ΔT) between the centre and outer parts of the reaction cell always exists and maximal ΔT of 150°C was estimated for the temperature of 1600-1700°C. The duration of isothermal heat treatment at a maximal temperature was varied from 15 to 60 min. The detailed assembling of high-pressure cell was reported previously [35]. To study the effect of the temperature (T), the duration isothermal heat treatment (τ), the heating and cooling rates (V) on crystal growth and superconducting properties of grown crystals various T-τ regimes were examined at the constant pressure 5 GPa.

Single crystals of $MgB_2$ were mechanically isolated from the bulk samples and examined by the powder X-ray diffractometer MXP–18 (MAC Science) with Ni-filtered $CuK_\alpha$ radiation and the four-circle X-ray diffractometer (AFC5R, Rigaku) with $MoK_\alpha$ radiation monochromized by graphite. Crystal morphology was studied by the optical microscope (Vanox-T, Olympus) and the scanning electron microscope (JSM-5400LV, JEOL). Superconducting properties of single crystals were studied by DC magnetization using SQUID (MPMS XL, Quantum Design) and four-probe resistivity measurements at magnetic fields up to 12 Tesla.

## Results and Discussions

Figure 1 shows the P-T reaction diagram for Mg-B-N system with the positions of important melting and decomposition lines plotted according to the results of previous studies. The melting temperature of Mg with pressure ($T_m$) was reported by Kennedy and Newton [36] up to 4.5 GPa and it was calculated for a higher pressure range [37] using relation: $T_m=T^o_m [1+C(\Delta V/V_o)]$, where $T^o_m$ is the melting temperature of Mg under ambient pressure; $(\Delta V/V_o)$ is the isothermal dilatation and C is the constant determined from the data of Kennedy and Newton. The pressure dependence of the melting temperature for two metastable eutectics $Mg_3BN_3$–hBN ($T_{e1}$) and $Mg_3B_2N_4$-hBN ($T_{e2}$) was reported by Gladkaya e.a. [38] and Endo e.a. [37]. The pressure dependence of of $MgB_2$ decomposition temperature ($T_{dec}$) in the "$MgB_2$-BN" system was estimated in our previous study [35]. Note that "$MgB_2$-BN system" indicates only the composition of the initial reaction mixture, since the $MgB_2$-BN quasibinary section of the ternary Mg-B-N system does not exist.

Figure 2 shows several typical regimes in the T-$\tau$ space: rapid heating at a rate of 100°C/min to the reaction temperature (1); step-heating with one or two isothermal stages at about 950 and 1300°C prior to the final heat treatment at the maximal temperature 1600-1650°C (2); quenching from the maximal to the room temperature (A); cooling at a constant rate of 3-5°C/min to a low temperature region with subsequent quenching (B). The filled areas at isothermal stages correspond to the maximal temperature difference within a reaction cell. The temperatures of Mg melting ($T_m$), melting of $Mg_3B_2N_4$-hBN ($T_{e1}$) and $Mg_3B_2N_4$-hBN eutectic ($T_{e2}$) and decomposition of $MgB_2$ ($T_{dec}$) at 5 GPa are indicated by dotted lines.

**(i) Crystal growth and effect of temperature and heating process on the size and shape of crystals**

Figure 3 shows the temperature dependence of the maximal size of forming crystals for 30 min. of isothermal heat treatment at 5 GPa. We found that the size of the $MgB_2$ crystallites slowly increases during heating in the wide temperature range of 950-1500°C. Quenching experiments showed the formation of almost single-phase $MgB_2$ and no visible traces of its interaction with BN. The crystal growth is drastically accelerated at T>1600°C. In this temperature range, partial decomposition of MgB$_2$ was observed in the central part of the reaction cell. This process was accompanied by the formation of the complicated phase mixture containing MgB$_4$ and intermediate products of Mg interaction with the BN (Mg$_3$BN$_3$, Mg$_3$B$_2$N$_4$). According to the reference reaction diagram (Fig.1), the latter phases could form the binary eutectic liquid with the hexagonal BN (hBN). Moreover, the formation of more complicated ternary eutectics (e.a. MgB$_4$-Mg$_3$BN$_3$-hBN) is also possible [39]. Note that formation of the eutectic liquid leads to dissolution of the BN reaction cell at the isothermal heat treatment stage. For the wall thickness of 1 mm, the maximal duration of the isothermal heat treatment at 1600$^o$C was about 1 hour and it reduces to 30 min at T=1650$^o$C. It was reported that oxides impurities in the BN (especially B$_2$O$_3$) can strongly affect on the formation of the eutectic liquid at high pressure and temperature [40,41] by the following reactions:

Mg$_3$B$_2$N$_4$ +3/2 O$_2$ =3 MgO +2BN +N$_2$ or  Mg$_3$B$_2$N$_4$+3O$_2$ =Mg$_3$(BO$_3$)$_2$ +2N$_2$.

Indeed, our experiments showed that utilization of high-purity BN containers (N-1) reduces the amount of MgO impurity phase in comparison with the low-purity BN (HC) though both types of containers can be successfully utilized for crystal growth of MgB$_2$.

Owing to the temperature gradient, the MgB$_2$ remains stable in the colder part of the reaction cell. The coexistence of MgB$_2$ with the liquid promotes re-crystallization process and formation of larger crystals at the interface of MgB$_2$ and liquid phase. The inset of Fig.3 shows the scanning electron microscope pictures of crystals with typical morphologies, mechanically extracted from the MgB$_2$ bulk sample. It was found that rapid heating (regime 1) is more favourable for the formation of needle-like crystals (top left picture) and thin plate-like crystals with a rectangular shape (bottom left), while step-heating (regime 2) promoted growth of crystals with hexagonal-like shape (top right) and rather thick bar-like crystals (bottom right). The shortest dimension always corresponds to the c-axis of the crystals, which reflects the anisotropic nature of $MgB_2$. The typical dimension of the crystals that can be extracted from the bulk was about 300-500 μm in the (ab)-plane and the maximal size was about 700 μm. A few crystals with the size of about 1 mm were extracted from some bulk samples although their formation is not reproducible. The typical thickness was 50-70 μm for the crystals grown by regime 1, and it was almost doubled for the crystals grown by regime 2.

The observed variation of the $MgB_2$ crystal shape can be qualitatively explained by the variation of temperature difference in the reaction cell. Rapid heating of un-reacted mixture of reagents (regime 1) to a high-T region (1600-1650°C) leads to a very fast formation and growth of $MgB_2$ crystallites. It is well known that fast growing rate in addition to a large temperature gradient is favourable for the formation of crystal of non-equilibrium shapes (e.a. whiskers or thin plates) due to the substantial growing instabilities, which was actually observed in our experiments. Preliminary heat treatment at a low-T range (950-1300°C), where the temperature gradient is not so large and the growing rate is low, is more favourable for the formation of small isotropic crystallites. Further re-crystallization of these particles at a high-T range does not change much the initial morphology of crystals and rather thick plate- or bar-like crystals can be grown (regime 2).

These results show that the temperature and the heating regimes of high-pressure experiments in Mg-B-N system allow to control the size and morphology of growing $MgB_2$ single crystals, which must be quite important, since for the majority of physical measurements appropriate size and shapes of crystals are very crucial.

Our preliminary experiments showed that variation of cooling rate does not change the crystal size. In the following section, we emphasize on the effect of cooling rates on the structure and the physical properties of $MgB_2$ single crystals.

**(ii) Effect of cooling rate on the structure and the physical properties**

The typical temperature dependent magnetization was measured on the bulk $MgB_2$ samples isolated from the high-pressure cell with an applied dc field of 10 G. The zero-field (ZFC) and field-cooled (FC) magnetization data are shown in Fig.4. The inset shows the expanded view of the magnetization near $T_c$ for the samples cooled by regimes A and B. For different batches of samples A, the variation of the $T_c(onset)$ value was in the range of 37.7-38.1 K, while samples B reproducibly showed the slightly higher value $T_c(onset)=38.4$-38.7 K.

To clarify the origin of this difference in $T_c$, several tens of the crystals were mechanically extracted from the bulk samples A and B. The crystals were aligned at the surface of sample holder and attached by vacuum grease. Figure 5 shows the typical X-ray diffraction patterns for the aligned single crystals. Very sharp peaks corresponding to (00l) reflections can be seen. The inset shows the typical line profiles of (002) peak for crystals A and B. The line width for crystals A is larger than that for crystal B, though the full width at half maximum value (FWHM) of 0.1 deg is significantly smaller than the values reported for $MgB_2$ powder samples [11,13]. For the best crystals B the FWHM value was close to the instrumental resolution (0.05 degree) determined by measurements of standard samples – high quality MgO and $Al_2O_3$ single crystals.

The larger broadening of peaks for crystals A than for crystals B suggests a higher concentration of defects in the crystals quenched from the high temperature in comparison with the slowly cooled ones. It is well known that X-ray line broadening can be caused by small crystallite or domain size (D<1000-2000Å) and lattice defects. For single crystal samples, the former source can be excluded and the difference in the peak width can be assigned to the effect of lattice strain. The same conclusion was made, based on the analysis of X-ray line broadening of $MgB_2$ powdered samples and two possible origins of the lattice strains were proposed. One is the presence of MgO inclusions inside MgB$_2$ grains [11] and the other is Mg-vacancies [13]. Although in both of those studies the linear correlation between the lattice strain and T$_c$ was found, the estimated strain value ($\varepsilon$=10$^{-2}$-10$^{-3}$) seems to be too large. The $\varepsilon$ calculated from our single crystal data is of order 10$^{-4}$. Therefore, we can speculate that the effect of small particle size cannot be neglected for interpretation of the peaks broadening of MgB$_2$ powder samples. As to the origin the observed differences in the FWHM for our crystals, we think that it is not due to a difference in concentration of MgO inclusions, because cooling rate cannot strongly affect the concentration of MgO inclusions in the MgB$_2$ crystals. In contrast, lattice strains originated from the presence of Mg-vacancies can be a much more plausible candidate for the lattice defects. Our structure refinement analysis for the MgB$_2$ single crystals showed the probability of the 4-5% Mg-deficiency in the crystals grown under high pressure [19], which is in good agreement with the value of 3-5% reported for the powder samples with T$_c$=36.6-37.8 K [13].

In this study, 5 single crystals from each series of crystals A and B were selected for structure analyses. The lattice constants for these 10 crystals are shown in Table 1. The variation of lattice constants can be seen for both series. The $c/a$ ratio shows less fluctuation within series of 5 crystals and average value $c/a$ =1.1410(2) was estimated for crystals A and 1.1405(1) for crystals B. It was reported that Mg-deficiency increases $c$ and decreases $a$ lattice constant [11,13,15,16] and minimal $c/a$ value corresponds to the stoichiometric MgB$_2$. Therefore, we can speculate the smaller Mg-deficiency for crystals B than for crystals A. However, the refinement of the Mg-occupancy showed the variation of the Mg content in the range from Mg$_{0.941(5)}$B$_2$ to Mg$_{0.962(4)}$B$_2$ for the both series crystals. It means that the difference in stoichiometry of the crystals, if any, is too small to be reliably detected in our structural experiments. Low temperature studies for crystals shaped to a sphere, to minimize the thermal parameters and uncertainty of the absorption correction, are in progress. A more precise structural refinement of the Mg-occupancy can be helpful to solve the contradictions in the reported experimental data for the dependence of $T_c$ on Mg-deficiency [11-18] and the pressure dependence of $T_c$ ($dT_c/dP$) obtained for polycrystalline samples with different stoichiometry [42-45].

The origin of $T_c$ difference in our crystals can be explained by variation of lattice constants. It was reported that monotonic decrease of $T_c$ with pressure arise predominantly from the decrease of electron-phonon coupling parameter $\lambda$ due to lattice stiffening, and not from electronic effects [44]. Since compression decreases both lattice constant $a$ and $c$ it is not yet known which one of them is responsible for decreasing $T_c$. Our single crystal data show that larger $a$ parameters can be responsible for higher $T_c$ value for crystals B in comparison with crystals A, while $c$ parameters were quite similar for both crystals.

The crystals with different $T_c$ were used for further studies to clarify the effect of lattice defects on the superconducting properties of $MgB_2$. Figure 6 (a) shows the magnetic hysteresis curves M(H) at 10 K for 100 crystals A and B mechanically aligned at the surface of sample holder with H//c orientation. The field dependence of critical current density at 10 and 20 K (Fig.6b) was estimated by the Bean model using formula $J_c=20\Delta M/d$. Here the volume magnetization was calculated from the theoretical density of $MgB_2$ $\rho=2.628 g/cm^3$, where the unit cell volume was determined from single crystal XRD data and the crystals weight was measured by the microbalance with an accuracy of 10μg. Based on the optical microscopy observation, the average crystal size d=250 μm was estimated for both samples. The $J_c$ values estimated at zero magnetic field shows a small difference (about 20%) between crystals A and B. Crystals A show slightly higher $J_c$ value, while the difference become negligible even at a low magnetic field. These data showed that the lattice strain induced by Mg-deficiency or Mg-vacancies themself do not produce the strong pinning centres in the bulk crystals. The values of $J_c < 10^5$ A/cm$^2$ (at T=10 and 20 K) are more than an order of magnitude lower than those reported for polycrystalline samples [2]. The low value of $J_c$ and its fast suppression by low magnetic fields (H<3000 Oe) suggest a high perfection of our crystals.

The temperature dependent resistivity for two crystals selected from different batches A and B are presented in Fig.7. The resistivity for both crystals was normalized at 300 K to emphasize the difference in the residual resistivity ratio (RRR). The RRR($R_{300K}/R_{40K}$) =5.15 was estimated for crystal A and 7.25 for crystal B. The inset shows the expanded view of resistivity near the $T_c$. For both crystals superconducting transitions are very sharp $\Delta T_c$ <0.3 K and crystal A has the lower $T_c$(zero)=37.7K than the $T_c$(zero)=38.4K for crystal B. The absolute value of resistivity at 40 K estimated for several crystals A was about $m\Omega$/cm and it was apparently smaller for crystals B.

At present there is no consensus about the value of RRR and residual resistivity intrinsic for highly pure MgB$_2$. The $\rho$(40K) as low as 0.4 $m\Omega$/cm and RRR as high as 25 were reported for the polycrystalline samples with $T_c$ of 39 K [17]. It was shown that purity of boron reagent, nominal composition of the sintered samples, un-reacted impurity (e.a. Mg) and processing parameters can strongly affect $\rho$(40K) and RRR data [18]. The results presented in the ref.[18] showed that polycrystalline samples with the high $T_c$=39.2K and the large RRR>15 can be synthesized only from the isotopically pure crystalline $^{11}$B (99.95%), whereas utilization of more pure amorphous boron (99.99%) gives the lower $T_c$(zero)=38.6K and smaller RRR=7.0 that are essentially the same as our data for crystal B. The origin of the observed difference between amorphous and crystalline B reagents is not clear yet. From this point of view it is important to analyse the recent data on the high-quality MgB$_2$ film fabricated on the surface of 99.99% pure B crystal that showed the highest reported $T_c$ (onset)=41.7K and $T_c$(zero)=39.7K [46]. For such a film the RRR=7.5 and $\rho(40K)=0.7 m\Omega$/cm was estimated. The authors suggest that the tensile strain from the boron crystal substrate that changed the lattice constants of $MgB_2$ may play an important role for the observed increase of $T_c$ value.

Finally we discuss the difference in the critical magnetic field between crystals A and B based, on the study of their transport properties. The resistive transitions for crystal A and B at various magnetic fields for H//c are shown in Fig.8. The differences between two samples can be clearly seen at a magnetic field higher than 2Tesla where $T_c$ decreases more rapidly for crystal B. The observed difference shows that crystal A has a higher value of second critical field in comparison with the crystal B. For H//ab orientation (fig.9) the differences in the magnetic field behaviour of resistive transition are negligible for the magnetic field H<8 Tesla, but at a higher field region some differences can be detected for the $T_c$(onset).

The observed differences in critical current density, residual resistivity and upper critical fields between crystals are in good agreement with the higher concentration of lattice defects in crystals A in comparison with crystals B.

Despite of the observed differences, the value of electronic anisotropy ($\gamma$) for upper critical field ($H_{c2}$) was quite similar for both crystals. At T=20 K, the $\gamma$ value close to 3.0 was estimated from the onset of resistive transitions and it was about 4.2 if $H_{c2}(T)$ line was defined by the temperatures where R(T)=0. In some of the recent reports it was claimed that electrical transport measurements are not well suited to probe the bulk upper critical filed $H_{c2}(T)$ [9,10]. It was proposed that for H//c the resistive onset is a manifestation of the onset of surface superconductivity on the vertical side faces of the plate-like crystals [9,26]. More detailed transport and magnetization measurements of samples reported in the present study will be presented elsewhere [47].

## Conclusion

Our experiments show that Mg-B-N system is very promising for crystal growth of $MgB_2$ under high pressure. Formation of various binary and ternary eutectic liquids at a high pressure and temperature as well as increased stability of $MgB_2$ phase, give an opportunity to accelerate crystal growth and to provide crystals with a dimension of several hundreds microns for less than 1 hour. Size and morphology of $MgB_2$ single crystals can by controlled by temperature and thermal gradients in the heating process. The grown single crystals have obvious advantages in the purity, size and variety of crystal shapes, compared with the crystals grown by heat treatment in the sealed metal containers. By cooling slowly from the maximal temperature, the concentration of lattice defects and/or the coherent inner strains reduced, which affects various properties of $MgB_2$.

## Acknowledgements

This work is supported by the New Energy and Industrial Technology Development Organization (NEDO) as Collaborative Research and Development of Fundamental Technologies for Superconductivity Applications.

# Figure Captions

1. Pressure dependence of Mg melting temperature ($T_m$) [36,37], melting temperatures of $Mg_3BN_3$–hBN ($T_{e1}$) [37,38] and $Mg_3B_2N_4$–hBN eutectics ($T_{e2}$) [38] and decomposition temperature of $MgB_2$ ($T_{dec}$) [35]

2. Typical regimes in temperature-time (T-$\tau$) space utilized for crystal growth of $MgB_2$ at 5 GPa. The black regions at isothermal stages correspond to the temperature difference between central and outer part of reaction cell. Dotted lines indicate the temperature of Mg melting ($T_m$), melting of eutectic in Mg-B-N system ($T_{e1}$ and $T_{e2}$) and thermal decomposition of $MgB_2$ ($T_{dec}$) at 5 GPa.

3. Temperature dependence of maximal size of $MgB_2$ crystals for 30 min of isothermal heat treatment at 5 GPa. Inset shows SEM pictures of extracted crystals of different morphology.

4. A typical temperature dependence of the magnetization after zero field cooling (ZFC) and field cooling (FC) in a magnetic field 10G for bulk sample sintered at P=5GPa, T=1650$^{o}$C for 30 minutes. Inset: expanded view of temperature dependences of magnetization near $T_c$ for different bulk samples A and B.

5. XRD patterns of mechanically aligned $MgB_2$ single crystals. Inset shows the typical line profiles of the 002 peak for crystals A (dashed-dotted line) and B (dotted line). Solid line shows the sharpest line for crystals B to illustrate the instrumental resolution. The lines have been normalized to emphasize the difference in the line width.

6. a) The magnetic hysteresis loops (M-H) at 10 K for crystals A (open square) and B (solid square) for H//c; b) Magnetic field dependence of critical current density for crystals A (open symbols) and B (solid symbols) measured at 10 and 20 K for H//c.

7. Temperature dependence of the normalized resistivity for single crystal A (solid) RRR=5.15 and B (open) RRR=7.25. Inset: expanded view of superconducting transition near $T_c$

8. Superconducting transitions at various magnetic fields applied parallel to the c-axis (H//c) of: a) crystal A and b) crystal B.

9. Superconducting transitions at various magnetic fields applied parallel to the ab-plane (H//ab) of: a) crystal A and b) crystal B.

Table1. Lattice parameters, superconducting transition temperature (T$_c$), residual resistivity ratio (RRR) and full width at half maximum (FWHM) of 002 peak for crystals A and B

| Crystal | $a$ (Å) | $c$ (Å) | c/a | T$_c$ (K) | RRR (R$_{300K}$/R$_{40K}$) | FWHM [002] (deg) |
|---|---|---|---|---|---|---|
| **A** 1 | 3.0868(5) | 3.5215(5) | 1.1408 | 37.7-38.1 | 5.5±0.5 | 0.1±0.02 |
| 2 | 3.0860(11) | 3.5210(12) | 1.1410 | | | |
| 3 | 3.0861(8) | 3.5209(5) | 1.1409 | | | |
| 4 | 3.0852(8) | 3.5202(8) | 1.1410 | | | |
| 5 | 3.0857(12) | 3.5211(17) | 1.1411 | | | |
| **B** 1 | 3.0873(9) | 3.5208(7) | 1.1404 | 38.4-38.7 | 7.5±0.5 | 0.06±0.01 |
| 2 | 3.0874(7) | 3.5213(6) | 1.1405 | | | |
| 3 | 3.0869(4) | 3.5205(5) | 1.1405 | | | |
| 4 | 3.0877(5) | 3.5214(6) | 1.1405 | | | |
| 5 | 3.0871(5) | 3.5210(6) | 1.1406 | | | |

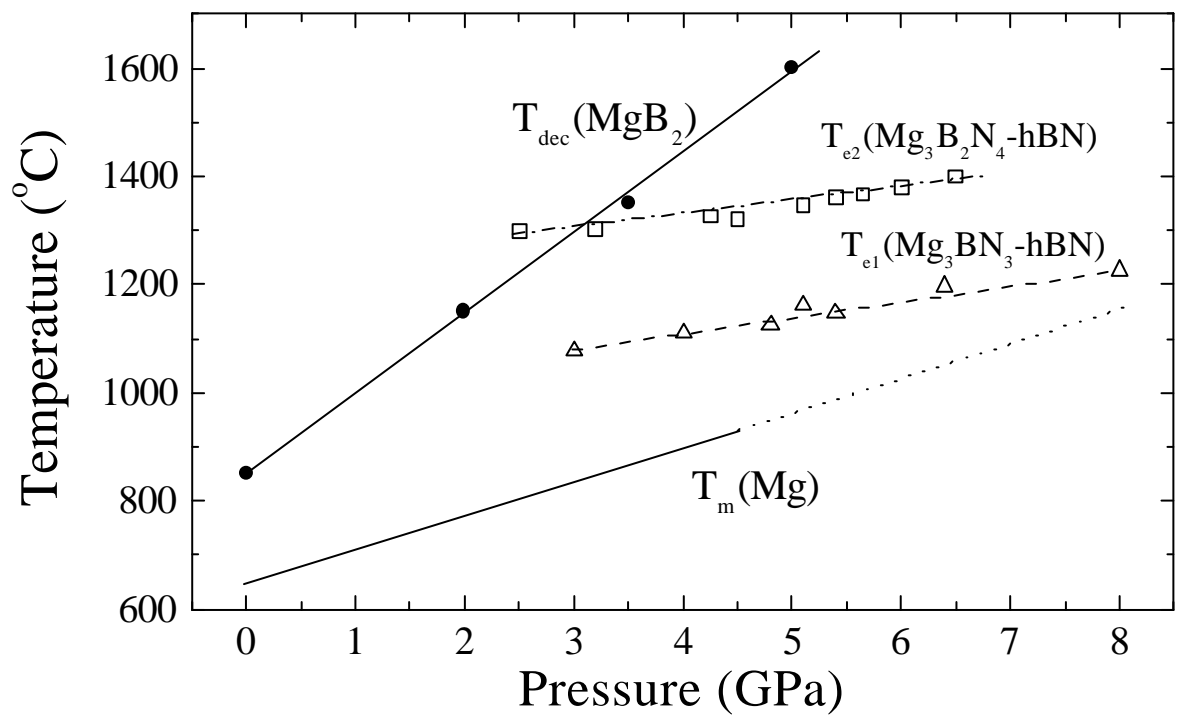

Fig.1 (S.Lee e.a.)

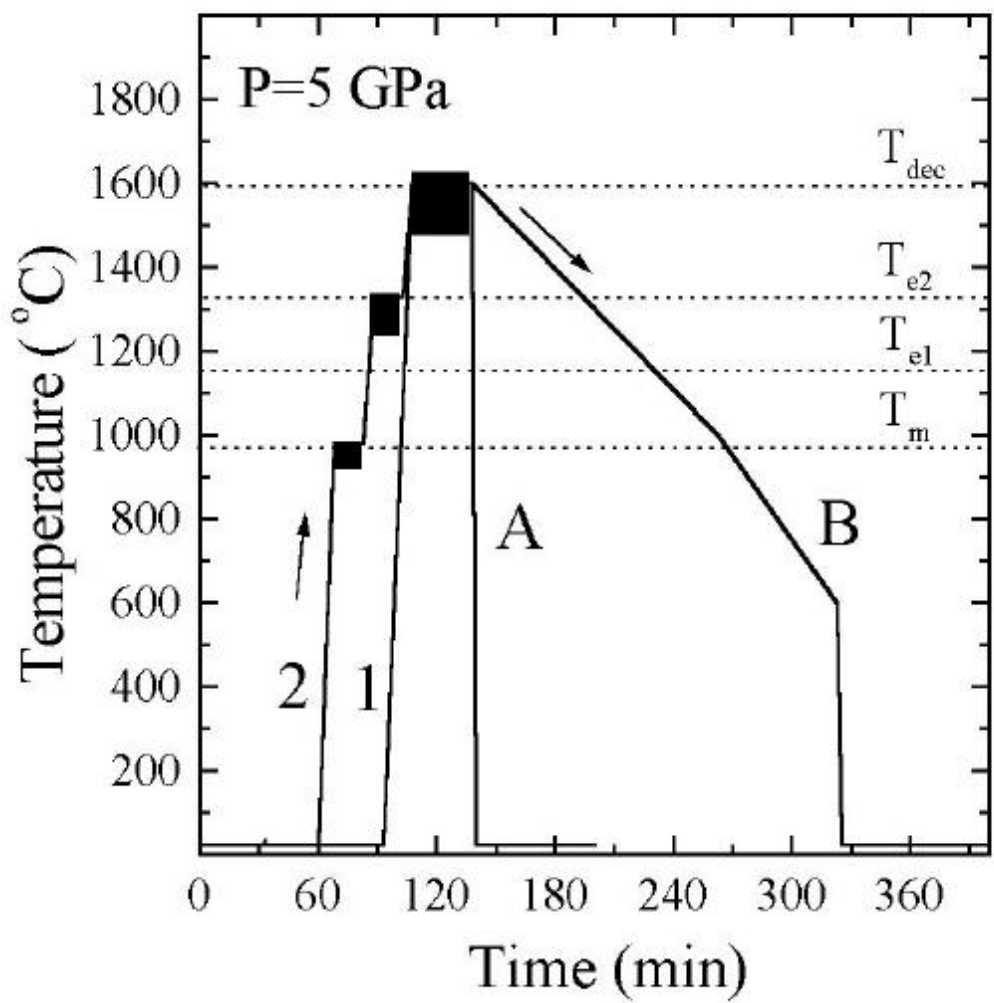

Fig.2 (S.Lee e.a.)

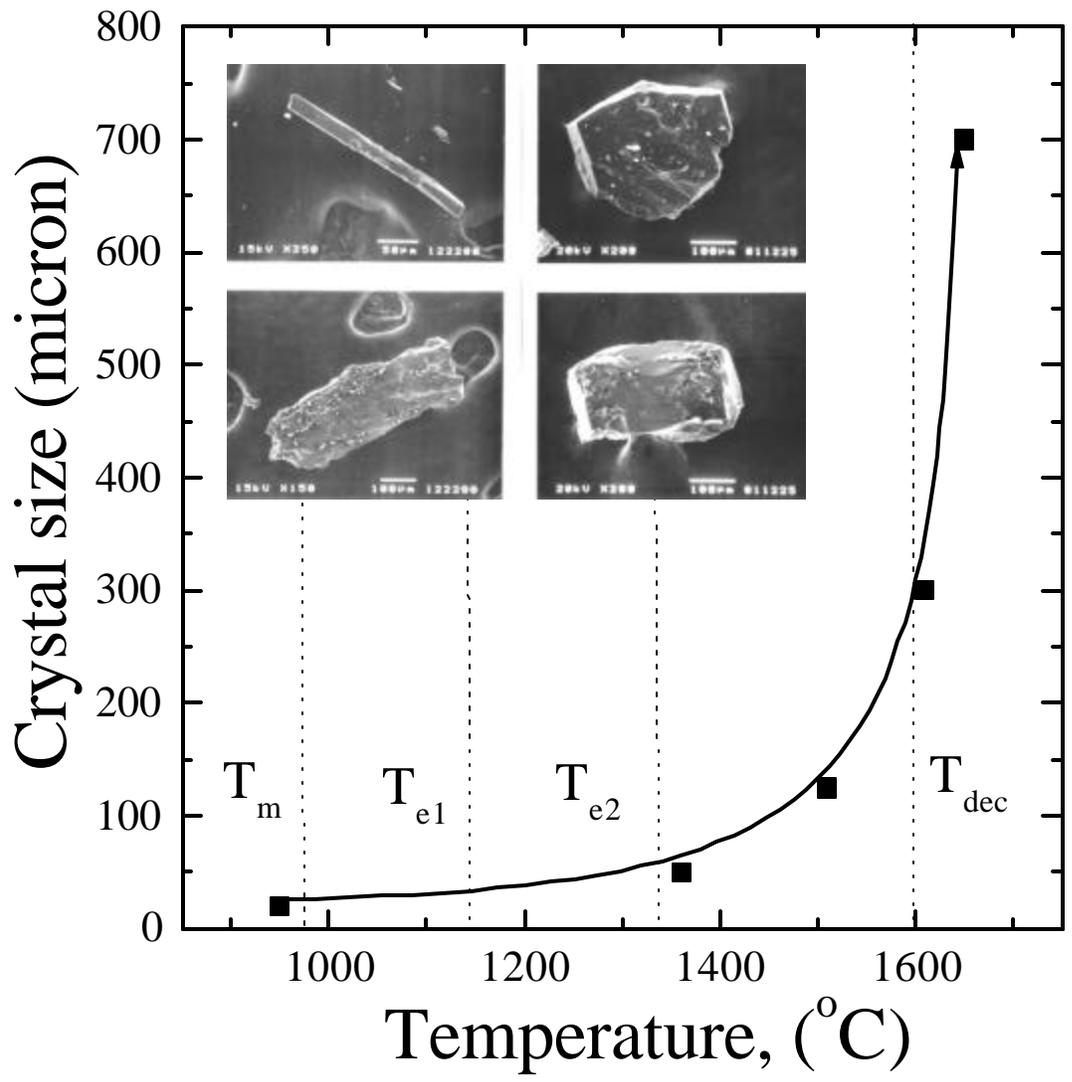

Fig.3 (S.Lee e.a.)

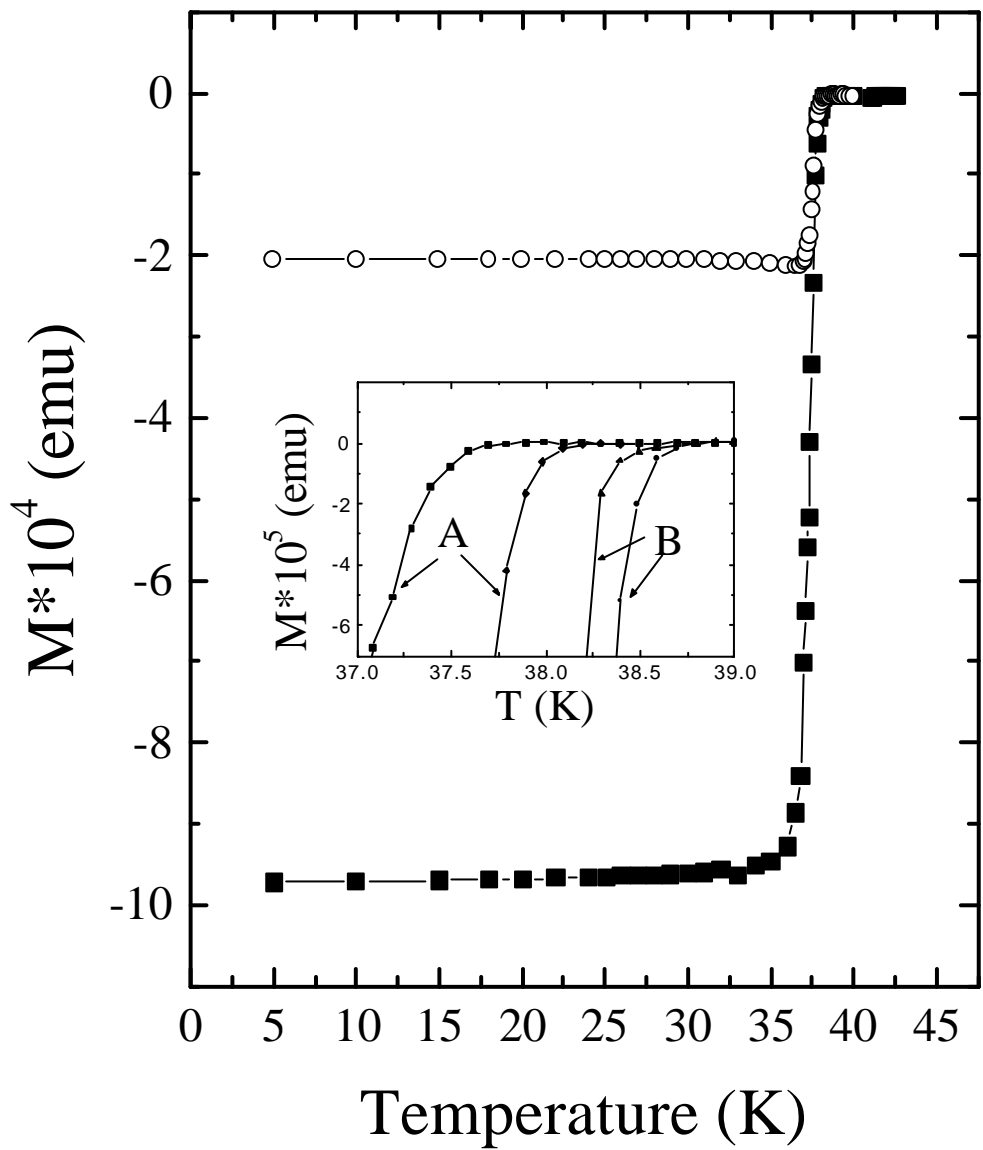

Fig.4 (S.Lee e.a.)

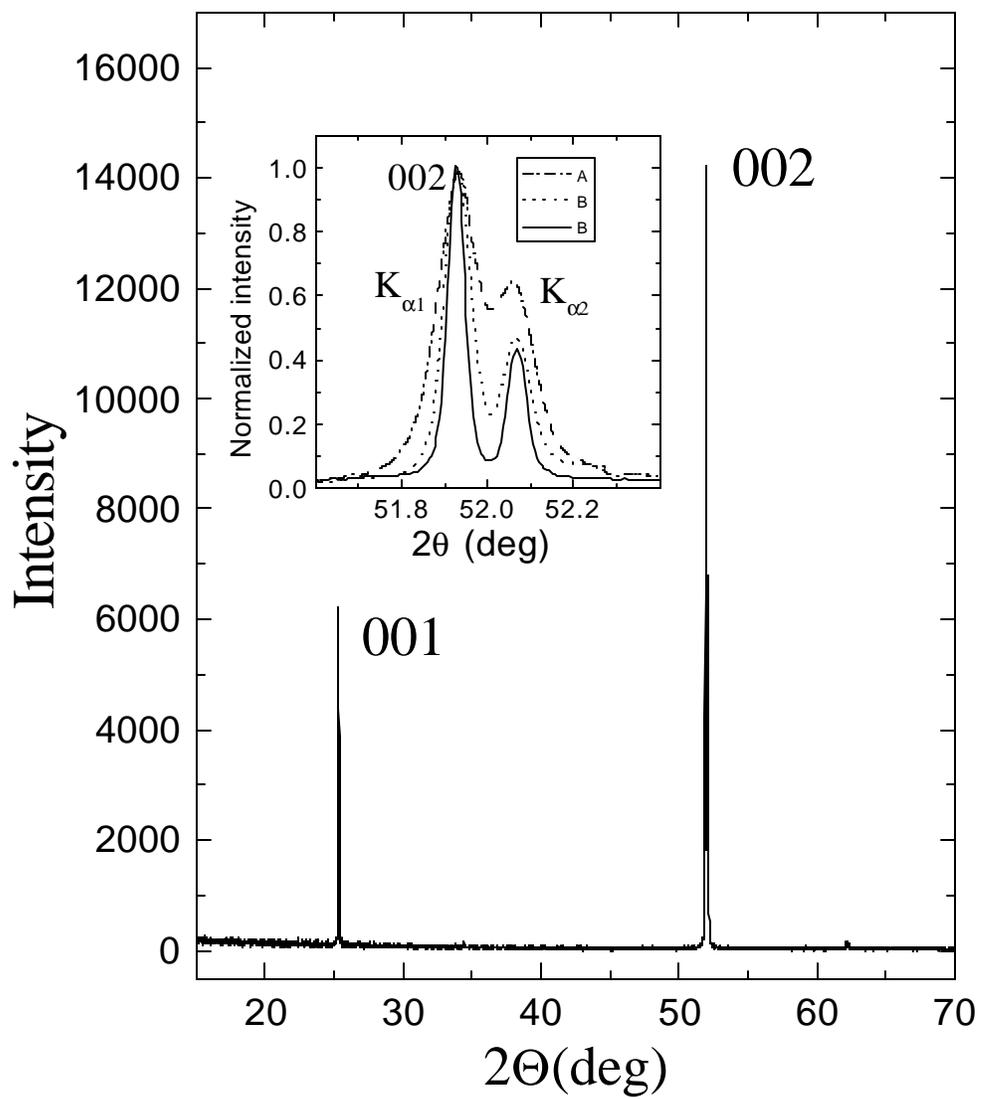

Fig.5 (S.Lee e.a.)

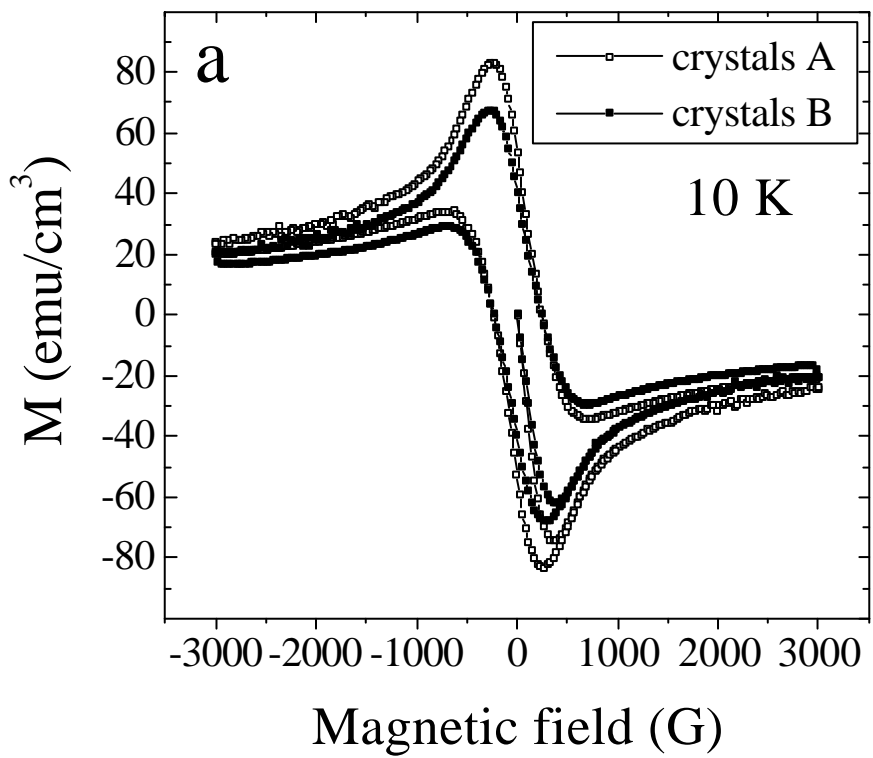

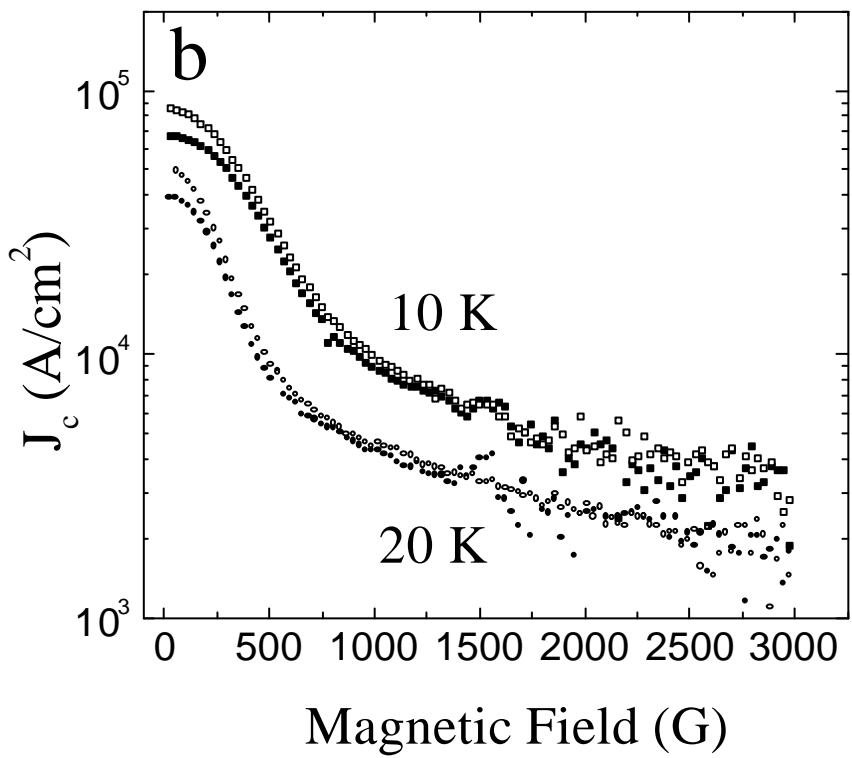

Fig.6 (S.Lee e.a.)

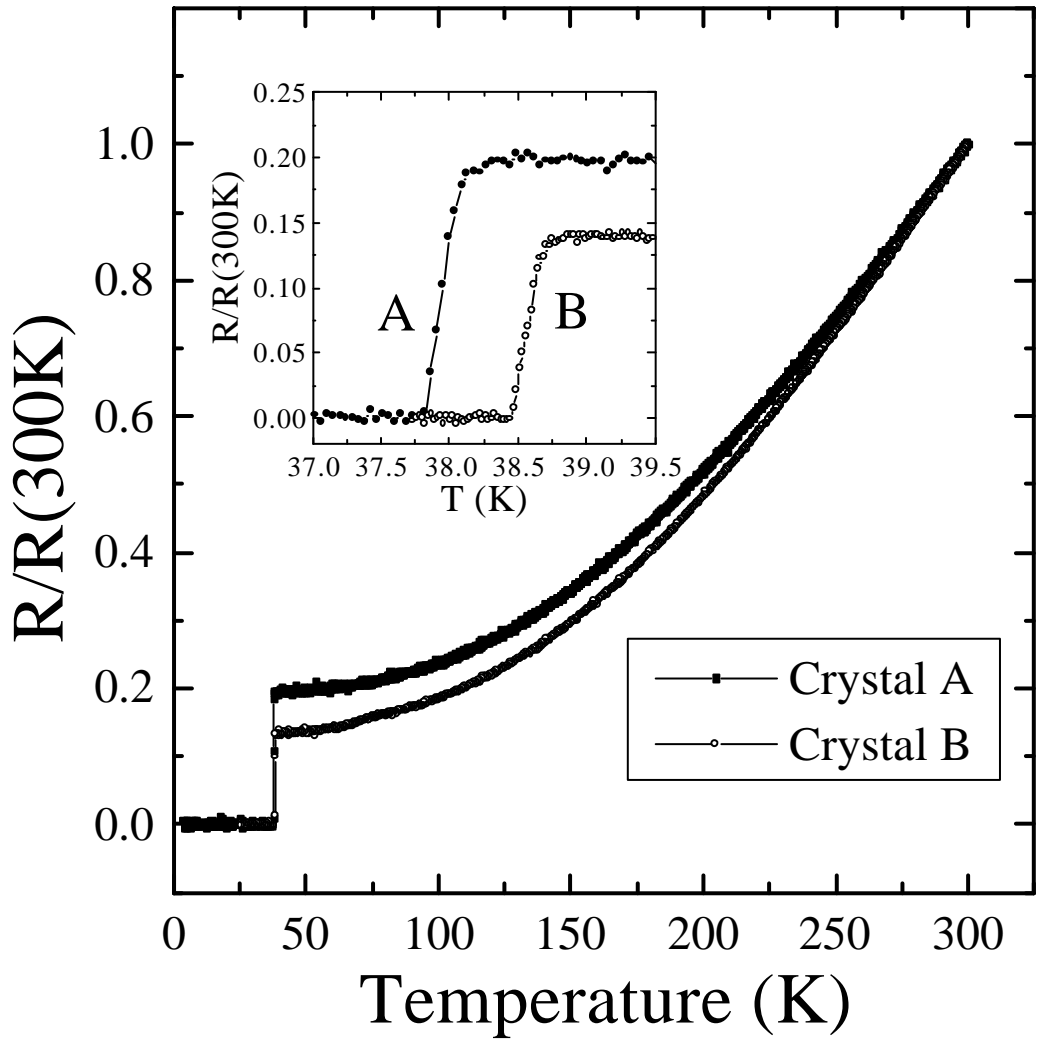

Fig.7 (S.Lee e.a.)

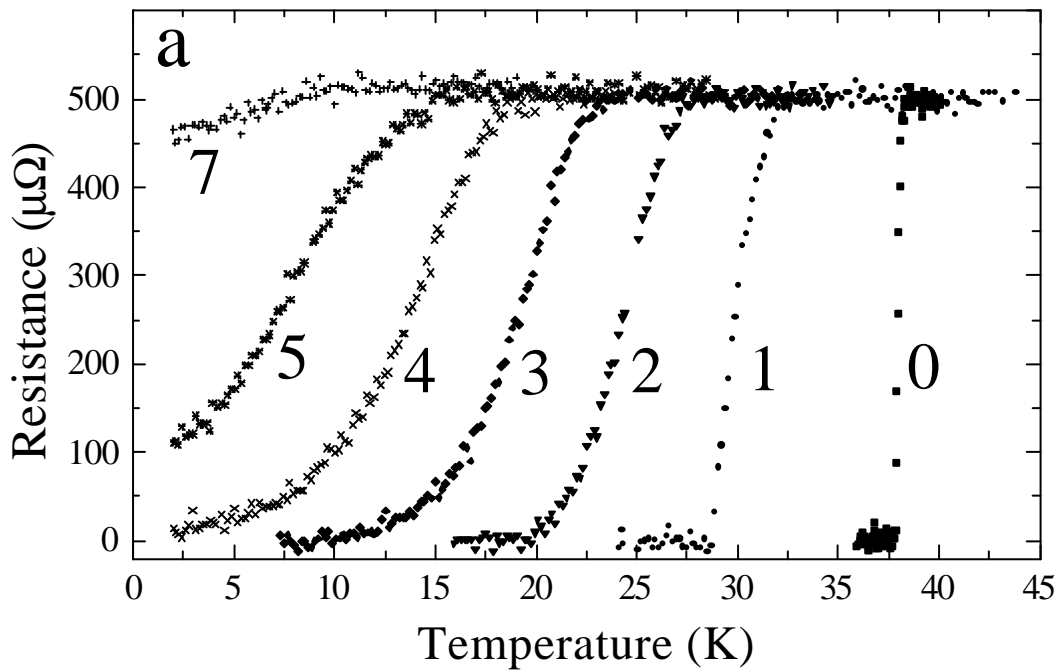

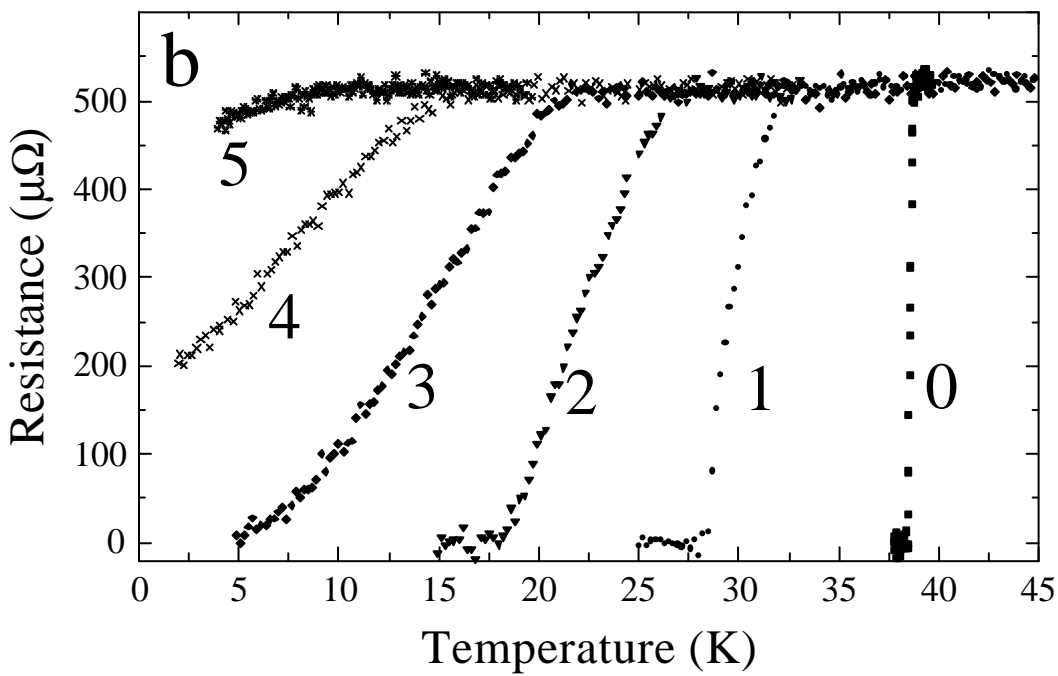

Fig.8 (S.Lee e.a.)

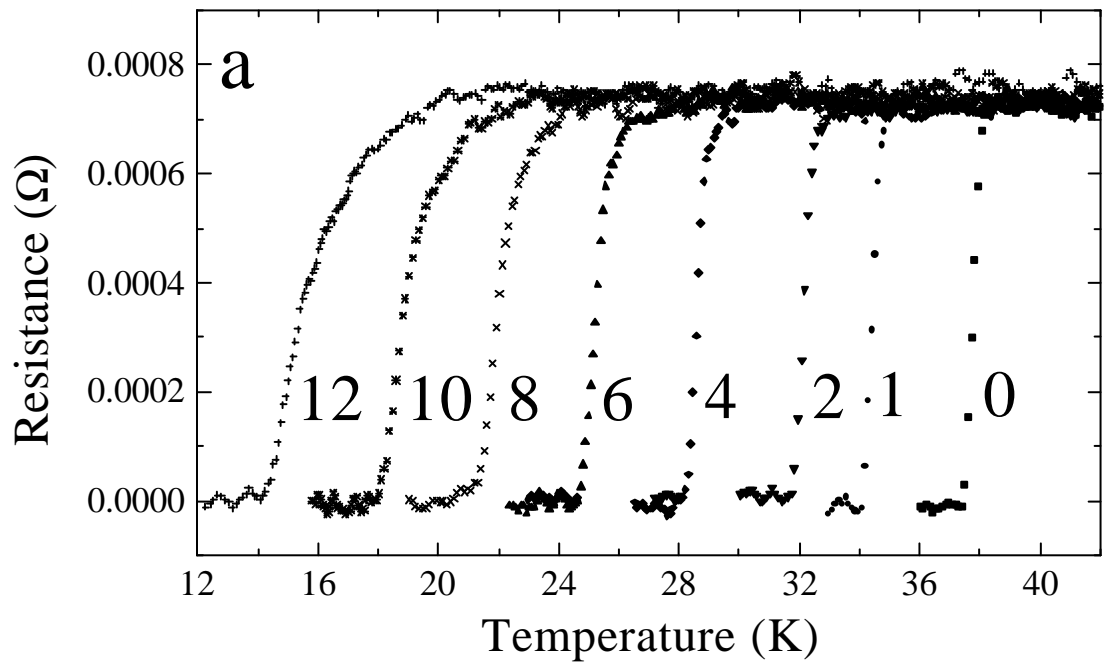

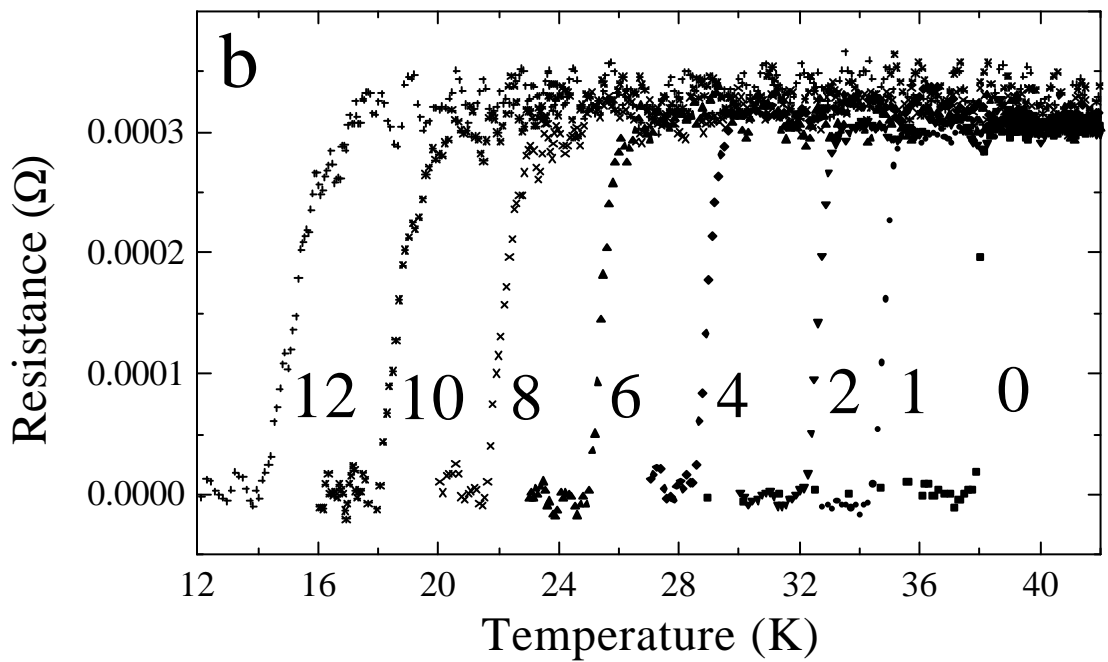

Fig.9 (S.Lee e.a.)